\documentclass[aps,prd,twocolumn,groupedaddress,showpacs]{revtex4-2}

\usepackage{graphicx}% Include figure files
\usepackage{color}%for preprints
\usepackage[T1]{fontenc}
\usepackage{amssymb,amsmath}
\usepackage{ulem}

\newcommand{\bfv}{\mbox{\boldmath $v$}}

\newcommand{\bfp}{\mbox{\boldmath $p$}}
\newcommand{\bfq}{\mbox{\boldmath $q$}}

\newcommand{\bfB}{\mbox{\boldmath $B$}}

\newcommand{\bfh}{\mbox{\boldmath $h$}}

\begin{document}

%\preprint{}

\title{ Distribution function of nuclei from  $e^\pm$ scattering in the presence of a strong primordial magnetic field}


\author{Motohiko Kusakabe}
\email{kusakabe@buaa.edu.cn}
\affiliation{
  School of Physics, and
  International Research Center for Big-Bang Cosmology and Element Genesis,
  Beihang University 37, Xueyuan Rd., Haidian-qu, Beijing 100083 China}

\author{Atul Kedia}
\affiliation{
  Center for Astrophysics, Department of Physics, University of Notre Dame, Notre Dame, IN 46556, U.S.A.}

\author{Grant J. Mathews}
\affiliation{
  Center for Astrophysics, Department of Physics, University of Notre Dame, Notre Dame, IN 46556, U.S.A.}

\author{Nishanth Sasankan}
\affiliation{
  Center for Astrophysics, Department of Physics, University of Notre Dame, Notre Dame, IN 46556, U.S.A.}

%\thanks{}


\date{\today}

\begin{abstract}
  The amplitude of the primordial magnetic field (PMF) is constrained from observational limits on primordial nuclear abundances. Within this constraint, it is possible that nuclear motion is regulated by Coulomb scattering with electrons and positrons ($e^\pm$'s), while $e^\pm$'s are affected by a PMF rather than collisions. For example, at a temperature of $10^9$ K, thermal nuclei typically experience $\sim 10^{21}$ scatterings per second that are dominated by very small angle scattering leading to minuscule changes in the nuclear kinetic energy of order $\mathcal{O}$(1) eV. In this paper the upper limit on the effects of a possible discretization of the $e^\pm$ momenta by the PMF on the nuclear momentum distribution is estimated under the extreme  assumptions that the momentum of the $e^\pm$ is relaxed before and after Coulomb scattering to Landau levels, and that during  Coulomb scattering the PMF is neglected. This assumption explicitly breaks the time reversal invariance of Coulomb scattering, and the Maxwell-Boltzmann distribution is not a trivial steady state solution of the Boltzmann equation under these assumptions. We numerically evaluate the collision terms in the Boltzmann equation, and show that the introduction of a special direction in the $e^\pm$ distribution by the PMF generates no directional dependence of the collisional destruction term of nuclei.
  Large anisotropies in the nuclear distribution function are then constrained from big bang nucleosynthesis.  Ultimately, we conclude  that a PMF does not significantly affect  the isotropy or BBN.
\end{abstract}

%\pacs{26.35.+c, 04.50.Kd, 98.80.Es, 98.80.Ft}
%26.35.+c Big Bang nucleosynthesis
%04.50.Kd Modified theories of gravity
%98.80.Es Observational cosmology (including Hubble constant,
%          distance scale, cosmological constant, early Universe, etc)
%98.80.Ft Origin, formation, and abundances of the elements (see also 26.35.+c Big Bang nucleosynthesis Nuclear astrophysics)

\maketitle


\section{Introduction}\label{sec1}
In the early universe, nuclei are strongly coupled with electrons and positrons ($e^\pm$) via Coulomb scattering \cite{Voronchev:2008zz,Sasankan:2019oee}.  A standard assumption during big bang nucleosynthesis (BBN) is that the nuclear energy distribution obeys  Maxwell-Boltzmann (MB) statistics as a nonrelativistic approximation to the Fermi-Dirac (FD) and Bose-Einstein (BE) distribution functions \cite{Hayashi1956,Wagoner:1966pv}.

Simulations of nuclear motion in a relativistic $e^\pm$ plasma during  the BBN epoch have been recently performed with Coulomb scatterings taken into account.  It was shown that the rapid thermalization of nuclei leads to  an MB distribution \cite{Sasankan:2019oee,Kedia:2020xdc}.
A perturbative analysis of the relativistic Boltzmann equation shows that any fractional deviation of the  thermal nuclear distribution from the MB distribution is of the order $10^{-18}$ at a temperature of $T =1$ GK, limited by the factor of $H/\Gamma_{\rm Coul}$, where $H$ is the cosmic expansion rate and $\Gamma_{\rm Coul}$ is the scattering rate of nuclei \cite{McDermott:2018uqm}.

If there is a nonthermal particle source such as decaying exotic particles and evaporating black holes in the early universe, high energy tails in the spectra of photons and nuclei can be deformed and the BBN yields could be affected \cite{1979MNRAS.188P..15L,1988PhRvD..37.3441R,1988ApJ...330..545D,Kedia:2020xdc}. Also,  the effect of replacing an MB distribution for nuclei with Tsallis distributions \cite{Tsallis:1987eu} has been studied.  It was shown that this can change the nuclear reaction rates and primordial abundances \cite{Bertulani:2012sv,Hou:2017uap}. Note, however, that if the nuclear distribution functions differ from MB statistics, then the relative velocity distribution depends upon the  masses of reacting nuclei in contrast to the case of MB in which it only depends upon the reduced mass \cite{Kusakabe:2018dzx}. In addition, if the MB statistics is modified  to Tsallis statistics, the phase space is changed from the simple product of the distribution functions of reacting particles as prescribed in the standard reaction theory \cite{Rueter:2019ubf}.

A possible source of deviation of particle distributions from MB is the existence of a primordial magnetic field (PMF).
A strong magnetic field changes weak \cite{1969Natur.222..649O,Cheng:1996yi,Luo:2020slj} and nuclear \cite{Kawasaki:2012va} reaction rates, as well as thermodynamic properties of the $e^\pm$s \cite{Kernan:1995bz,Kawasaki:2012va} through Landau discretization of the particle momenta. In addition, the PMF energy enhances the cosmic expansion rate \cite{1969Natur.223..938G}, and mainly through this effect the PMF is constrained to be $B_0 < 1.5 \mu$G at present or $B< 2.0 \times 10^{11}$G at $T_9 \equiv T/(10^9~{\rm K}) =1$ \cite{Kawasaki:2012va,Luo:2018nth}.
  The BBN constrains PMFs generated earlier in the big bang, i.e., the electroweak phase transition \cite{Vachaspati:2020blt}, neutrino decoupling \cite{Dolgov:2001nv}, and so on.
These small-scale PMFs, however, undergo dissipation during the later cosmic evolution \cite{Brandenburg:2003pe,Banerjee:2004df,Widrow:2011hs,Vachaspati:2020blt}. This PMF evolution depends upon the helicity and homogeneity of the magnetic fields \cite{Brandenburg:2003pe,2020PhRvD.102b3536B}, and its behavior in the late non-linear structure formation phase has been investigated in a number of recent three-dimensional simulations \cite{Hutschenreuter:2018vkr,Sanati:2020oay,Vazza:2020phq,Katz:2021iou}. Ultimately, what we can observe with current measurements are signatures of large-scale PMFs which could have been generated in the inflationary epoch before the big bang \cite{1992ApJ...391L...1R}.
Constraints on the PMF \cite{Widrow:2011hs} have been deduced from various astronomical observations such as the cosmic microwave background (CMB) temperature and polarization from gravitational \cite{Zucca:2016iur,BICEP2:2017lpa,Sutton:2017jgr,Pogosian:2018vfr,Yamazaki:2018gmr,Paoletti:2018uic,Minoda:2020bod} and heating effects \cite{Paoletti:2019pdi}, the redshifted 21-cm signal from heating effects \cite{Minoda:2018gxj,Natwariya:2020ksr,Katz:2021iou,Bera:2020jsg}, ultra-faint dwarf galaxies from gravitational effects \cite{Safarzadeh:2019kyq}, and gravitational waves (at very small scales) \cite{Saga:2018ont}.

 It has been pointed out that an inhomogeneous PMF with a comoving amplitude of $B_0 ={\mathcal O}(0.1)$ nG and a coherence length of $\lambda ={\mathcal O}(10)$ kpc can induce chemical separations of Li$^+$ ions from gravitationally collapsing cosmic structures \cite{Kusakabe:2014dta,2019ApJ...876L..30K}. As a result, this sub-nG magnetic field could explain the long-standing discrepancy between the Li abundances in metal-poor stars \cite{Spite:1982dd,Sbordone:2010zi} and the standard BBN prediction (see Ref. \cite{Hayakawa:2021jxf} for the latest result based upon recent experiments on the $^7$Be($n$, $p$)$^7$Li reaction).
 Interestingly a PMF with $B_0 ={\mathcal O}(0.1)$ nG over similar scales also induces baryon inhomogeneities before the epoch of cosmological recombination \cite{Jedamzik:2020krr}, and its effect on the CMB could also help explain the difference between Hubble parameters deduced from observations of Type Ia supernovae and the CMB \cite{Reid:2019tiq}.

 The magnetic field amplitude is also constrained from the CMB power spectra and baryon acoustic oscillation data \cite{Aghanim:2018eyx}. The constraint on the number of effective relativistic degrees of freedom is $N_\mathrm{eff} =2.99 \pm 0.17$. This corresponds to $B_0^2 =(-0.13 \pm 0.27)~\mu$G$^2$ in terms of energy density, based upon the relation:
 \begin{eqnarray}
 B_0^2 & \approx& \frac{7 \pi^3}{15} \left(\frac{4}{11} \right)^{4/3}
 \left(N_\mathrm{eff} -N_\mathrm{eff,0} \right) T_0^4 \nonumber \\
 &=& 2.383 \left(N_\mathrm{eff} -N_\mathrm{eff,0} \right) \mu \mathrm{G}^2,
 \label{correspondence1}
 \end{eqnarray}
where $N_\mathrm{eff,0} =3.046$ is the contribution from three generations of relic neutrinos and antineutrinos, and $T_0 =2.7255$ K is the present CMB temperature. The $2 \sigma$ upper limit on the comoving field strength is then $B_0 < 0.82~\mu$G. However, we caution the reader regarding three points here: (1) since most of the small-scale fields would dissipate during a long cosmic time from the BBN epoch to the time of CMB last scattering, this limit can not be compared simply with the BBN constraint. (2) The PMF affects BBN not only via the cosmic expansion rate but also changes in $e^\pm$ distribution function. Therefore, the correspondence of Eq. (\ref{correspondence1}) is only an approximation even in the BBN epoch. Hence, a consistent calculation including various aspects of PMF effects \cite{Kawasaki:2012va} is necessary. (3) A rigorous calculation of the CMB including PMF effects is also required as suggested in Ref. \cite{Jedamzik:2020krr}.

In this letter, we show details of the collision term in the Boltzmann equation for nuclei during BBN and calculate the possible effects of a PMF on the nuclear distribution and resultant light-element abundances  produced during  BBN.
First, it is  shown that,  under existing constraints on the  possible amplitudes for the PMF, nuclei are predominantly affected by Coulomb scattering with $e^\pm$s, the momenta of which are regulated by the PMF if the field strength is as high as $\approx 10^{11}$G.
Second, from a numerical integration of the collision term from Coulomb scattering, it is seen that nuclei experience very frequent scatterings most of which result in energy shifts by only ${\mathcal O}(10^{-5})T$.
Third, the effects on the nuclear distribution function from the  introduction of a discretization of $e^\pm$ momenta is analyzed.  It is confirmed that the  Coulomb scattering does not lead to a significant change in the nuclear isotropy and energy distributions.
Finally, using a toy model, we illustrate  the  possible effects of a hypothetical anisotropy in the nuclear distribution function on the light elements produced during BBN.

We note that the evaluation of the collision term in the Boltzmann equation, performed here, does not depend on whether the magnetic field is homogeneous or not, on all but the smallest scales. As noted below electrons and positrons significantly change their directions on very short time scales. This means that the mean free path is very short $O(10^{-7})$ cm. The corresponding comoving scale is only $O(10^2)$ cm. At each position of a nucleus, only the magnetic field amplitude over this small scale matters. If there is an inhomogeneity in the magnetic field over this or smaller scales, the formulation here may need modification. However, the current formulation is applicable for inhomogeneity over larger scales. The purpose of the current letter is the first evaluation of detailed statistical properties of the effect of Coulomb scattering on the nuclear distribution function. We then show that for a typical upper limit on the magnetic field amplitude during BBN, the effects of magnetic fields via changes in the electron and positron distribution functions are small enough. This means that even in a universe with a spatially inhomogeneous field strength, any deviation in the nuclear distribution from a Maxwell-Boltzmann form at any location can be safely neglected.

%
%Finally, we summarize this study.

\section{Timescales}\label{sec2}
%\paragraph{Time scales}
First, we show that during the BBN epoch at $T_9 =1$, the motions of $e^\pm$s are controlled by the magnetic field if the field is strong, i.e., $B \sim 10^{11}$G, while the motion of nuclei is governed by Coulomb collisions with $e^\pm$s.

\subparagraph{$e^\pm$:}
The gyrofrequency is a typical rate for the magnetic field effect on charged particles \cite{1975clel.book.....J}. For $e^\pm$ it is given by
\begin{eqnarray}
  \omega_g &=&eB/E \nonumber \\
  &=& 3.52 \times 10^{18}~{\rm s}^{-1}
  \left( \frac{B}{10^{11}~{\rm G}} \right) \left( \frac{E}{0.511~{\rm MeV}} \right)^{-1},
  \label{eq2}
  %2.e11 *5.9157e-21/ 0.511e-3 /6.582119514e-25 =3.51762446857544e+18
  %e =alpha^0.5
  %  =0.085424543
  %alpha =1./137.035999139
  %1 e G =5.9157e-21 GeV^2
  %1 =6.582119514e-25 GeV s
\end{eqnarray}
where $E$ is the energy of the $e^\pm$.
Momentum exchange is dominated by Coulomb scatterings with abundant $e^\pm$s during BBN.
The momentum transfer cross section is estimated as follows. As a rough approximation we  consider scatterings off of target $e^\pm$ particles at rest in the frame of the cosmic fluid under the nonrelativistic approximation.
The maximum impact parameter is set to the Debye length given by
\begin{eqnarray}
  b_{\rm max} &=&\lambda_{\rm D} = 7.67 \times 10^{-9}~{\rm cm}
  ~({\rm at}~T_9 = 1)
\end{eqnarray}
with
\begin{eqnarray}
  \lambda_{\rm D} &=&\sqrt{T/(4 \pi n_{e} e^2)} \\
  &\approx & \sqrt{\frac{\pi^{1/2}}{2^{5/2} \alpha m_e^{3/2} T^{1/2}}}
    e^{m_e /(2 T)}
    ~({\rm for}~T \lesssim m_e),
  %(pi**0.5 /(2.**2.5 *alpha *0.5109989461e-3**1.5 *8.6173303e-5**0.5))**0.5 * exp(0.5109989461e-3 /(2.*8.6173303e-5))
  % =388079.08166 GeV^-1
  % =7.6736041055828e-09 cm
  %alpha =1./137.035999139
  %m_e =0.5109989461 MeV
  %1 =8.6173303e-5 GeV /GK
  %1 GeV^-1 =1.97733e-14 cm
\end{eqnarray}
where
$n_e =n_{e-} +n_{e+}$ is the total number density of $e^\pm$, $e$ and $m_e$ are the electronic charge and mass, respectively, and
$\alpha=e^2$ is the fine structure constant.

We use the relation between the scattering angle and the impact parameter,
\begin{equation}
  b_{\rm max} =\frac{q_1 q_2 e^2}{m v^2} \cot
  \left( \frac{\theta_{\rm min}}{2} \right),
  \label{eq6}
\end{equation}
where
$q_1$ and $q_2$ are charge numbers of reacting particles 1 and 2, respectively,
$m=m_e/2$ is the reduced mass,
$v$ is the $e^\pm$ velocity, and
$\theta_{\rm min}$ is the minimum scattering angle.
Then, the maximum of $\mu \equiv \cos{\theta}$ at $T_9=1$ is given by
\begin{eqnarray}
  \mu_{\rm max} &=& \cos \theta_{\rm min} \nonumber \\ %=
%  \cos \left[ 2 \cot^{-1} \left( b_{\rm max} \frac{m_e v^2}{2 \alpha} \right) \right] \\
%  &=& \cos \left[ 2 \cot^{-1} \left( 1.36 \times 10^{4} v^2 \right) \right]\\
%  &\approx & \cos \left[ \left( \frac{2}{1.36 \times 10^{4} v^2 } \right) \right] \\
  & \approx& 1 -1.08\times 10^{-8} v^{-4},
  %388079.08166 *0.5109989461e-3 /(2. *alpha)
  % =13587.6675772
%(pi**0.5 /(2.**2.5 *alpha *0.5109989461e-3**1.5 *8.6173303e-5**0.5))**0.5 * exp(0.5109989461e-3 /(2.*8.6173303e-5))
  % =388079.08166 GeV^-1
  %(2. /1.358e4)**2 /2. =1.08450224600415e-08
  %alpha =1./137.035999139
  %m_e =0.5109989461 MeV
\end{eqnarray}
where a small angle approximation was adopted.
With this maximum cosine $\mu_{\rm max}$, the momentum transfer cross section is given by
\begin{eqnarray}
  \sigma_{\rm mt}(\mu<\mu_{\rm max}) &\approx &\int_{-1}^{\mu_{\rm max}} \frac{d \sigma}{d\mu} \frac{\Delta q_e}{q_e} d\mu ~~,
 \nonumber  \\
%  &=&
%  \frac{2^{1/2} \pi \alpha^2}{v^2 q_e^2}
%  \left[ 2 (1-\mu)^{-1/2} +v^2 ( 1-\mu)^{1/2} \right]_{-1}^{\mu_{\rm max}} \\
%  &\approx&
%  \frac{2^{1/2}\pi \alpha^2}{v^2 q_e^2}
%  \left[ 2 (1-\mu_{\rm max})^{-1/2} \right] \\
  &\approx &
  4.54/q_e^2,
  %2.**0.5 *pi*1./137.035999139**2*(2. *1.08450224600415e-08**(-0.5))
  %=4.5437067
  \label{eq10}
\end{eqnarray}
where
$q_e$ and $\Delta q_e$ are the initial momentum and the momentum transfer of $e^\pm$, respectively,
and
the differential Mott cross section is given by
\begin{eqnarray}
  \frac{d \sigma}{d\mu}(v,\mu) &=&
  \frac{2\pi \alpha^2}{v^2 {q_e}^2}
  \frac{1}{(1-\mu)^2}
  \left[ 1 -\frac{v^2}{2} ( 1-\mu) \right].
  \label{eq_sigma_tot4}
\end{eqnarray}
The momentum transfer rate is then given by
\begin{eqnarray}
  \Gamma_{\rm Coul}(\mu<\mu_{\rm max}) &\approx& n_{e} \sigma_{\rm mt}(\mu < \mu_{\rm max}) \nonumber \\
%  &\sim & \sqrt{\frac{2}{\pi^3}} \left(\frac{T}{m_e} \right)^{3/2} m_e
%  \exp \left( -m_e /T \right)
%  \left( \frac{4.54}{v^2} \right) \\
%  &= & \frac{8.96 \times 10^{20}~{\rm s}^{-1}}{v^2}
%  \left(\frac{T}{m_e} \right)^{3/2} 
%  \exp \left( -m_e /T \right) \\
  &= & 1.65 \times 10^{17}~{\rm s}^{-1}v^{-2}~({\rm at}~T_9 =1).~~~~~
  %2.**0.5 /pi**1.5 *0.5109989461e-3 *4.5437067 /6.582119514e-25
  % =8.95890062098846e+20
  %2.**0.5 /pi**1.5 *(8.6173303e-2 /0.5109989461)**1.5 *0.5109989461e-3 *exp(-0.5109989461/8.6173303e-2) *4.5437067 /6.582119514e-25
  % =1.64953522983001e+17
  %m_n =0.939566 GeV
  %m_e =0.5109989461 MeV
  %1 =6.582119514e-25 GeV s
  %1 =0.1973269788 GeV fm
  %1 =8.6173303e-5 GeV /GK
\end{eqnarray}

The ratio of this scattering rate to the gyrofrequency is given by
\begin{eqnarray}
  R &=&
  \frac{\Gamma_{\rm Coul}(\mu<\mu_{\rm max})}{\omega_g} \nonumber \\
%  &\sim & \frac{255}{v^2}
%  \left(\frac{T}{m_e} \right)^{3/2} 
%  \exp \left( -m_e /T \right)
%  \left( \frac{B}{10^{11}~{\rm G}} \right)^{-1} \\
  & \approx & \frac{4.69 \times 10^{-2}}{v^2}
  \left( \frac{B}{10^{11}~{\rm G}} \right)^{-1}
  ~({\rm at}~T_9 = 1).
  %8.95890062098846e+20 /3.51762446857544e+18 =254.6861
  %1.64953522983001e+17 /3.51762446857544e+18 =0.046893
\end{eqnarray}

For a large enough field strength of $B \sim 10^{11}$G and $v \sim 1$, the Coulomb scattering rate is significantly smaller than the gyrofrequency. Therefore, the perturbation due to Coulomb scattering from the  discretized momentum distribution would be small.

\subparagraph{Nuclei:}
Scatterings are dominated by the Coulomb scattering with $e^\pm$. 
The maximum $\mu$ at $T_9=1$ is derived from Eq. (\ref{eq6}) to be
\begin{eqnarray}
  \mu_{\rm max} &=& \cos \theta_{\rm min} %=
%  \cos \left[ 2 \cot^{-1} \left( b_{\rm max} \frac{m_e v^2}{Z \alpha} \right) \right] \\
%  &=& \cos \left[ 2 \cot^{-1} \left( 2.72 \times 10^{4} Z^{-1} v^2 \right) \right]\\
%  &\approx & \cos \left[ \left( \frac{2}{2.72 \times 10^{4} Z^{-1} v^2 } \right) \right] \\
   \approx 1 -2.70\times 10^{-9} Z^2 v^{-4},
  %388079.08166 *0.5109989461e-3 /alpha
  % =27175.335154
%(pi**0.5 /(2.**2.5 *alpha *0.5109989461e-3**1.5 *8.6173303e-5**0.5))**0.5 * exp(0.5109989461e-3 /(2.*8.6173303e-5))
  % =388079.08166 GeV^-1
  %(2. /2.72e4)**2 /2. =2.70328719723183e-09
  %alpha =1./137.035999139
  %m_e =0.5109989461 MeV
\end{eqnarray}
where
$Z$ is the atomic number of the nucleus.
The momentum transfer cross section is then given by
\begin{eqnarray}
  \sigma_{\rm mt}(\mu<\mu_{\rm max}) &\approx &\int_{-1}^{\mu_{\rm max}} \frac{d \sigma}{d\mu} \frac{\Delta q_e}{q_e} \frac{q_e}{q_A} d\mu
\nonumber  \\
%  &=&
%  \frac{2^{3/2} \pi \alpha^2}{v^2 q_e q_A}
%  \left[ 2 (1-\mu)^{-1/2} +v^2 ( 1-\mu)^{1/2} \right]_{-1}^{\mu_{\rm max}} \\
%  &\approx &
%  \frac{2^{3/2} \pi \alpha^2}{v^2 q_e q_A}
%  \left[ 2 (1-\mu_{\rm max})^{-1/2} \right] \\
  &\approx &
  18.2 /(Z q_e q_A),
  %2.**1.5 *pi*1./137.035999139**2*(2. *2.70e-9**(-0.5))
  %=18.21267
  \label{eq_sigma_tot6}
\end{eqnarray}
where
$q_A$ is the nuclear momentum.
Accordingly, the momentum transfer rate is given by
\begin{eqnarray}
  \Gamma_{\rm Coul}(\mu<\mu_{\rm max}) &\approx& n_{e} \sigma_{\rm mt}(\mu< \mu_{\rm max}) \nonumber \\
%  &\approx &
%  4 \left( \frac{m_eT}{2 \pi} \right)^{3/2} \exp \left( -m_e /T \right)
%   \left( \frac{18.2 }{Z q_e q_A} \right) \\
%  &= & \sqrt{\frac{2}{\pi^3}}
%  \frac{18.2}{Z v \sqrt{2 m_A E_A}}
%  m_e^2 \left(\frac{T}{m_e} \right)^{3/2} 
%  \exp \left( -m_e /T \right) \\
%  &= & \frac{18.2}{\pi^{3/2} Z v \sqrt{m_A E_A}}
%  m_e^2 \left(\frac{T}{m_e} \right)^{3/2} 
%  \exp \left( -m_e /T \right) \\
%  &= & \frac{1.18 \times 10^{20}~{\rm s}^{-1}}{Z v}
%  \left(\frac{m_A}{m_n} \right)^{-1/2}
%  \left(\frac{E_A}{3~{\rm GK}/2} \right)^{-1/2}
%  \left(\frac{T}{m_e} \right)^{3/2} 
%  \exp \left( -m_e /T \right) \\
  &= & \frac{2.17 \times 10^{16}~{\rm s}^{-1}}{Z A^{1/2} v}
  ~({\rm at}~T_9 =1),~~~~~
  %18.21267 /pi**1.5 *0.5109989461e-3**2 / (0.939566 *1.5 *8.6173303e-5)**0.5 /6.582119514e-25
  % =1.17740915687641e+20
  %1.17740915687641e+20 *(8.6173303e-2 /0.5109989461)**1.5 *exp(-0.5109989461/8.6173303e-2)
  % =2.16787524089959e+16
  %m_n =0.939566 GeV
  %m_e =0.5109989461 MeV
  %1 =6.582119514e-25 GeV s
  %1 =0.1973269788 GeV fm
  %1 =8.6173303e-5 GeV /GK
\end{eqnarray}
where
$A$ is the nuclear mass number.

The ratio of this scattering rate to the gyrofrequency [Eq. (\ref{eq2})] is given by
\begin{eqnarray}
  R &=&
  \frac{\Gamma_{\rm Coul}(\mu< \mu_{\rm max})}{\omega_g} \\
  &\sim & \frac{6.15 \times 10^4}{v}
  \frac{A^{1/2}}{Z^2}
  \left(\frac{E_A}{3~{\rm GK}/2} \right)^{-1/2}
  \left(\frac{T}{m_e} \right)^{3/2} \nonumber \\
  && \times \exp \left( -m_e /T \right)
  \left( \frac{B}{10^{11}~{\rm G}} \right)^{-1} \\
  &= & \frac{11.3}{v}
  \frac{A^{1/2}}{Z^2}
  \left( \frac{B}{10^{11}~{\rm G}} \right)^{-1}~({\rm at}~T_9 =1).
  %1.17740915687641e+20 /3.51762446857544e+18 /0.511e-3 *m_n
  % =61543.80281
  %61543.80281 *(8.6173303e-2 /0.5109989461)**1.5 *exp(-0.5109989461/8.6173303e-2)
  % =11.331599
  %m_n =0.939566 GeV
\end{eqnarray}
For a large enough field strength, $B \sim 10^{11}$G, the Coulomb scattering rate is large enough when $T \sim m_e$, but diminishes relative to the gyrofrequency at lower temperatures.
Since this ratio is larger than that of $e^\pm$ at $T_9 =1$, we can expect situations in which the distribution of $e^\pm$ is predominantly governed by the magnetic field while those of nuclei are determined by Coulomb collisions with $e^\pm$ in the BBN epoch.

\section{Model}\label{sec3}
%\subparagraph{collision term}
Under the assumption of a homogeneous distribution of nonrelativistic nuclei during BBN, the invariant Boltzmann equation for nuclei in the $e^\pm$ plasma has the form of
\begin{eqnarray}
  && \frac{\partial f_A}{\partial t}
  +e \left(\bfv \times \bfB \right)^i \frac{\partial f_A}{\partial p_A^i} \nonumber \\
  &=&
  \frac{m_A}{E_A} \frac{g_e}{(2 \pi \hbar)^3}
  \int d^3 q_e \frac{E_e^{\rm (lab)}}{E_e}
  \int d\Omega_{\rm nuc} v \frac{d\sigma}{d \Omega_{\rm nuc}}
  \nonumber \\
  &&\times
  \left[f_A' f_e' (1 \pm f_A) (1 -f_e) -f_A f_e (1 \pm f_A') (1 -f_e') \right],~~~~~~
\end{eqnarray}
%\begin{widetext}
%\begin{eqnarray}
%  \frac{\partial f_A}{\partial t}
%  +e \left(\bfv \times \bfB \right)^i \frac{\partial f_A}{\partial p_A^i}
%  &=&
%  \frac{m_A}{E_A} \frac{g_e}{(2 \pi \hbar)^3}
%  \int d^3 q_e \frac{E_e^{\rm (lab)}}{E_e}
%  \int d\Omega_{\rm nuc} v \frac{d\sigma}{d \Omega_{\rm nuc}}
%  %\nonumber \\
%  %&&~~~~~~~~
%  %\times
%  \left[f_A' f_e' (1 \pm f_A) (1 -f_e) -f_A f_e (1 \pm f_A') (1 -f_e') \right],
%\end{eqnarray}
%\end{widetext}
where
$g_e=4$ is the sum of the statistical degrees of freedom of $e^-$ and $e^+$,
$m_A$ is the nuclear mass,
$E_e$ is the electron energy,
$f_A$ and $f_A'$ are the distribution functions of nuclear momenta in the initial and final states, $\bfp_A$ and $\bfp_A'$, respectively, and
$f_e$ and $f_e'$ are the distribution functions of $e^\pm$ momenta in the initial and final states, $\bfq_e$ and $\bfq_e'$, respectively.
The relative velocity $v(\bfp_A, \bfq_e)$ is equal to the electron velocity in the nuclear rest frame.
The quantity $d\sigma/d\Omega_{\rm nuc}$ is the differential Mott cross section with $\Omega_{\rm nuc}$ the solid angle in the nuclear rest frame, and the $E_e^{\rm (lab)}$ is the electron energy in the nuclear rest frame.

We take the $z$-axis as the field direction in a small local domain. The second term on the left hand side is zero if the gradient of the distribution function is proportional to the momentum along a plane perpendicular to the $z$-axis, i.e., $\partial f_A /\partial p_{Ai} \propto p_{Ai}$ for $i=x$ and $y$. We assume that the second term is zero hereafter.
On the right hand side, the first and second terms correspond to the production and destruction, respectively, from Coulomb scattering.

When $f_e$ and $f_e'$ obey an FD distribution, the quantity in the square brackets becomes zero for FD or BE distributions for nucleus $A$. Therefore, an FD or BE distribution is a trivial steady-state solution of the Boltzmann equation, and at BBN temperatures, both of them are well approximated by an MB distribution, i.e.,
\begin{eqnarray}
  f_{\rm MB}(\bfp_A) &=& n_A \left( 2 \pi m_A T \right)^{-3/2}
  \exp \left[ - p_A^2 /\left(2m_A T \right) \right],~~~~~
\end{eqnarray}
where
$n_A$ is the number density of the nuclide $A$ and
$p_A =|\bfp_A|$.

\subsection{$B = 0$ case:}\label{sec3a}
When there is no magnetic field, $e^\pm$s are assumed to have an FD distribution. Then, the production and destruction terms exactly cancel with each other. The destruction term is described by
\begin{widetext}
\begin{eqnarray}
  \left. \frac{df_A}{dt} \right|_{\rm des}(p_A; 0)
  &= & -
  f_A( p_A)
  \int d q_e
  \frac{q_e^2}{E_e(q_e)}
  f_e(q_e)
  \int d\Xi
  E_e^{\rm (lab)}(p_A, q_e, \Xi)
  v \left( p_A, q_e, \Xi \right)
  \nonumber \\
  &&
  \times
  \int_{-1}^{\mu_{\rm max}(p_A,\Theta; q_e)} d\mu
  \frac{d\sigma(v,\mu)}{d\mu}
  \frac{1}{2\pi} \int_{0}^{2 \pi} d\phi
  \left\{ 1 - f_e'\left( q_e'\left[ p_A, q_e, \Xi, \mu, \phi \right] \right)
    \right\}~~,
\end{eqnarray}
\end{widetext}
where
$\mu_{\rm max}(p_A,\Theta; q_e)$ is the maximum $\mu$ value as a function of $p_A$ and the angle between $\bfp_A$ and $\bfq_e$, i.e., $\Theta$, for a fixed $q_e =|\bfq_e|$.
The parameter $\Xi =\cos \Theta$, and $\phi$ is the azimuthal angle at the scattering.
The term $[ 1 -f_e'(\bfq_e') ]$ is from the Pauli blocking of $e^\pm$ with $\bfq_e' =\bfh(\bfp_A, \bfq_e, \mu, \phi)$ the momentum vector of $e^\pm$ in the final state.
The quantity $q^{\rm nuc}_e(q_e, p_A, \Theta)$ is the electron momentum in the nuclear rest frame, which appears in the differential Mott cross section.
Note that when $\bfp_A$, $\bfq_e$, and $(\mu,\phi)$ are given, $\bfp_A'$ and $\bfq_e'$ are determined.

In the BBN epoch, the $e^\pm$ plasma has an FD distribution when there is no magnetic field, i.e., $f_{e\mp}(\bfq_e) =1/ \left\{ \exp[(E_e \mp \mu_e)/T] +1 \right\}$ with $E_e=\sqrt{m_e^2 +q_e^2}$ the total energy. Unless $T \ll m_e$, the electron chemical potential is negligible, and we use the same function $f_e(\bfq_e)$ for $e^\pm$s.

The upper limit on the integral, i.e., $\mu_{\rm max}(p_A,\Theta; q_e)$ appears because of the minimum scattering angle $\theta_{\rm min}$.
The minimum angle is set so that the Mott scattering is ineffective if the impact parameter is larger than the Debye screening length. The effective minimum scattering angle is then given by
\begin{eqnarray}
  \theta_{\rm min}
%  2 \tan^{-1} \frac{\alpha}{m_e v^2}
%  \sqrt{\frac{4 \pi n_e \alpha}{kT}} \\
%  &=&2 \tan^{-1} \frac{\alpha}{(m_e kT)^{1/2} v^2}
%  \sqrt{\frac{4 \pi \alpha n_e }{m_e}} \\
  &=&2 \tan^{-1} \left\{ \alpha \omega_{\rm p} /\left[(m_e T)^{1/2} v^2 \right]
  \right\},
  \label{eq27}
\end{eqnarray}
where $\omega_{\rm p} =4\pi \alpha n_e/m_e$ is the plasma frequency of the background $e^\pm$ plasma.
%\begin{equation}
%  \omega_{\rm p}^2 = \frac{4\pi \alpha n_e}{m_e},
%  \label{eq_omega_pl}
%\end{equation}

\subsection{$B\neq 0$ case:}\label{sec3b}

We estimate the maximum possible effect of the $B$-field on nuclear distribution functions assuming that $e^\pm$s in their initial states only have a discretized momentum distribution. An upper limit on the magnetic field strength is adopted from BBN studies with a PMF \cite{Kawasaki:2012va}.

%\subsection{regulation of initial state distribution}
We assume some mechanism which only modifies the initial state momentum distribution of $e^\pm$s, such as a PMF.
An extreme case is considered here that the initial state energy spectrum is completely discretized into Landau levels.
The Pauli blocking for $e^\pm$ is negligible since the initial momentum of the $e^\pm$ is discretized and the final momentum is continuous in the present application. Since most scatterings result in small $\Delta \bfq_e$ (see below), transitions to different $n$ states are less important.
Therefore, the Pauli blocking effect is neglected in this case.

For the case of a finite $B$-field, the destruction term in the Boltzmann equation is described  by
\begin{eqnarray}
  \left. \frac{df_A}{dt} \right|_{\rm des}(p_A; B)
  &=&
  - \frac{m_A}{E_A} \frac{g_e}{(2 \pi \hbar)^3}
  f_A \left( \bfp_A \right)
  e B 
  \int_{-\infty}^\infty d q_{e\parallel} \nonumber \\
  &&
  \times \int d \phi_2 \sum_{n=0}^\infty \frac{\left( 2 -\delta_{n0} \right)}{2}
  \frac{E_e^{\rm (lab)}(p_A, q_e, \Xi)}{E_e(q_e)} \nonumber \\
  && \times
  f_{en}\left( q_{e \parallel}; T \right)
  v \left( p_A, q_e, \Xi \right) \sigma_{\rm tot}(v) \\
  f_{en}\left( q_{e \parallel}; T \right) &=& 
  \left\{ 1+ \exp \left[ E_{n}(q_z)  /T\right] \right\}^{-1},
\end{eqnarray}
where
$q_e = ( q_{e\perp}^2 + q_{e \parallel}^2)^{1/2}$ with $q_{e \perp}^2 =2 n eB$ and 
$E_n = ( q_e^2 +m_e^2 )^{1/2}$ are the momentum and energy of the $n^{\rm th}$ Landau level, 
$\delta_{n0}$ is the Kronecker delta, and
$\sigma_{\rm tot}$ is the total cross section for Coulomb scattering.

In the coordinates of $\bfp_A =p_A(\sin \theta_1, 0, \cos \theta_1)$ and $\bfq_e = q_e(\sin \theta_2 \cos \phi_2, \sin \theta_2 \sin \phi_2, \cos \theta_2)$, it follows that
\begin{eqnarray}
  \Xi &=&
  \sin \theta_1 \sin \theta_2 \cos \phi_2 + \cos \theta_1 \cos \theta_2.
\end{eqnarray}

\section{Results}\label{sec3c}
%\subsection{differential scattering rate}\label{sec4a}

\begin{figure}[tbp]
\begin{center}
\includegraphics[width=\linewidth,clip]{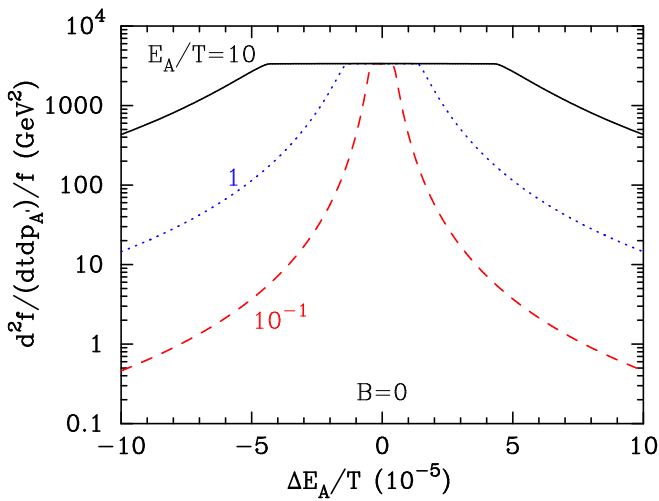}
\caption{Differential destruction term of the Boltzmann equation as a function of the change of the proton energy due to a scattering, for initial proton energies of $E_A/T=10^{-1}$, 1, and 10 at $T=10^9$ K in the case of no magnetic field.
  \label{fig1}}
\end{center}
\end{figure}

Fig. \ref{fig1} shows the differential rate of change of the distribution function $f(\bfp_A)$ with respect to the final state momentum $\bfp_A'$, i.e., $\partial^2 f(\bfp_A,\bfp_A')/(\partial t \partial \bfp_A')$, as a function of the change of the proton energy $\Delta E_A/T$ for  $p +e^\pm \rightarrow p + e^\pm$ scattering in the case of $B = 0$.
The central region of small $|\Delta E_A/T|$ has maximum rates since the small energy changes that result from small angle scatterings have  a huge cross section as long as the angle is above the lower limit. When the amplitude $| \Delta E_A/T |$ is larger, large angle scatterings are required. Since they occur with smaller differential cross sections, the rates are also smaller.

%*******************************************************

%*******************************************************

It is observed that the differential collision term has a plateau at $\sim 10^3 f$ GeV$^2$ in the region of $|\Delta E_A/T| \lesssim 10^{-5}$. 
The typical width of the distribution is rather consistent with the following analytical estimate.
The largest contribution to the collision integral comes from the parameter region with the smallest possible scattering angle for the thermal energy of the $e^\pm$. The minimum scattering angle for the thermal energy is given [Eq. (\ref{eq27})] by
\begin{eqnarray}
  \theta_{\rm min} &\approx
%  & 2 \frac{\alpha \omega_{\rm p}}{(m_e kT)^{1/2} v^2} \\
%  &=
  & 8.5 \times 10^{-5}
  \left( \frac{\omega_{\rm p}}{1.22 \times 10^{-6}~{\rm GeV}} \right)
  T_9^{-1/2} v^{-2}.~~~
  %8.51017e-5
\end{eqnarray}
For $p_A \gg q_e$ as in the case of thermal nuclei during the BBN epoch, we adopt an approximation that the Lorentz gamma for the nuclear velocity is $\gamma_A =(1-v_A^2)^{-1/2} \approx 1$, $q^{\rm nuc}_e \approx q_e$, and the angle of $\bfq^{\rm nuc}_e$ measured from the direction of $\bfp_A$, $\Theta' \approx \Theta$. Then it follows that
$\Delta p_A^2 \sim p_A q_e \theta$.
For this value, the change in the nuclear energy for $E_e/T \sim 1$ has a typical value of
\begin{eqnarray}
  \frac{|\Delta E_A|}{T}
  &\approx& 10^{-6} \sqrt{\frac{E_A}{T}} A^{-1}
  \left(\frac{\theta}{10^{-4}} \right)~~.
%  2.342178e-6
\end{eqnarray}

\begin{figure}[tbp]
\begin{center}
\includegraphics[width=\linewidth]{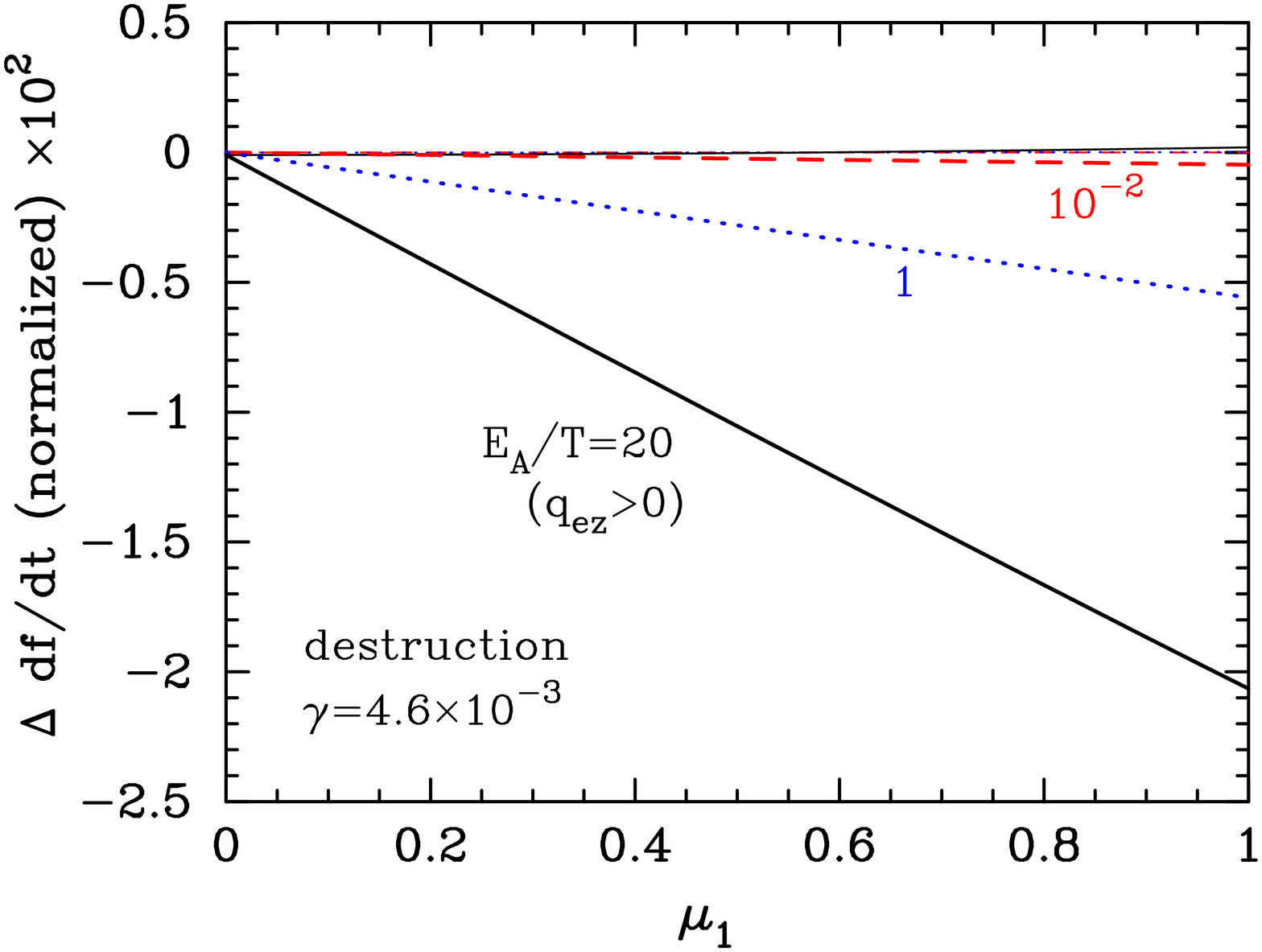}
\includegraphics[width=\linewidth]{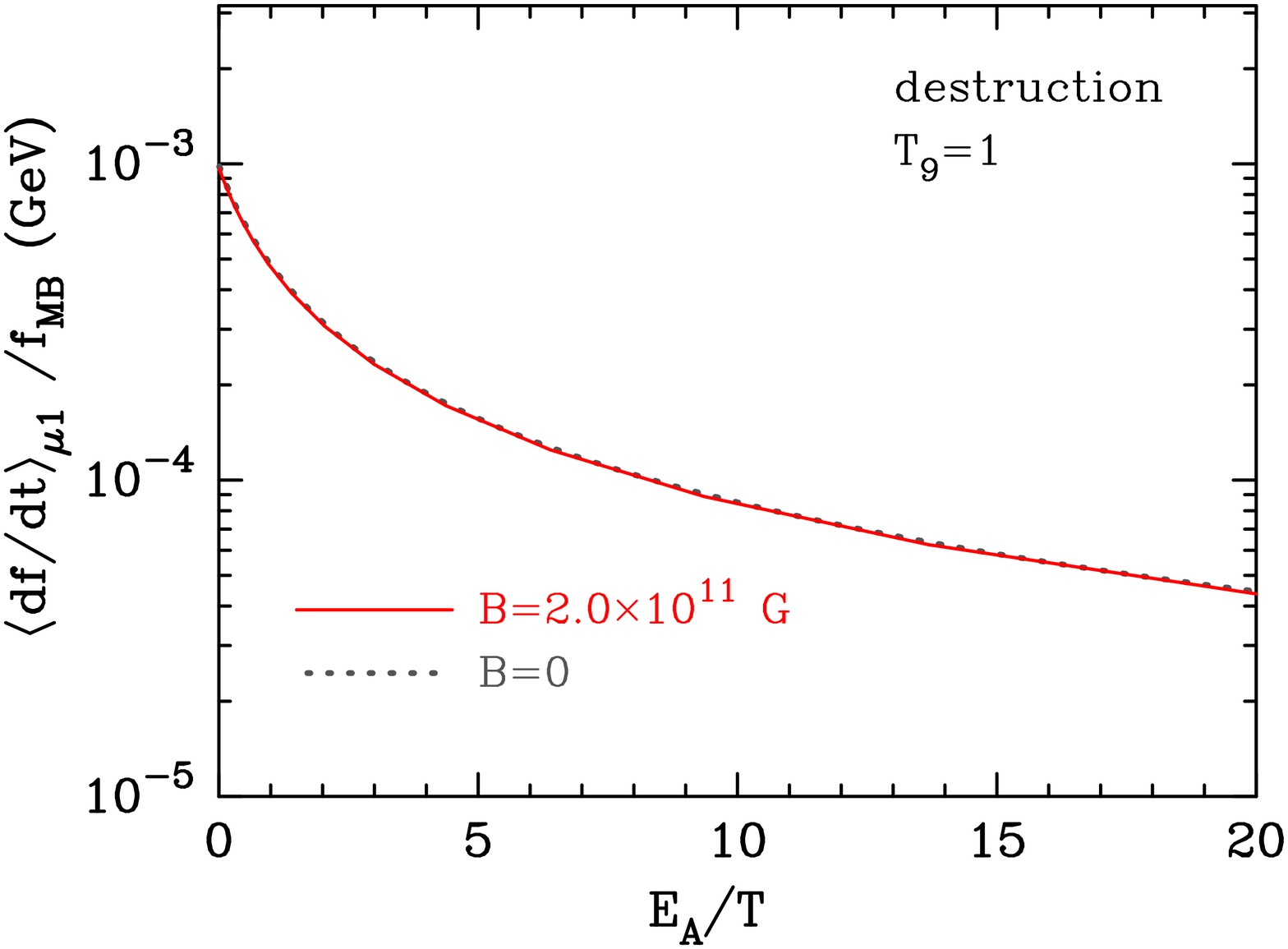}
\caption{(upper panel) Difference of the destruction term of the Boltzmann equation for protons as a function of the angle between the proton momentum and the magnetic field at $T=10^9$ K. The field is the largest allowed value, $2.0\times 10^{11}$ G.  Thin three lines show differences in the total destruction rate although they are nearly indistinguishable. The sloping three thick lines show differences in the partial destruction rate by $e^\pm$ particles with $q_{e\parallel}>0$;
  (lower panel) The destruction term averaged over the direction of the proton as a function of proton energy at $T=10^9$ K. Solid and dotted lines correspond to the results for $B=2.0 \times 10^{11}$ G and 0, respectively.
  \label{fig2}}
\end{center}
\end{figure}

The upper panel in Fig.~\ref{fig2} shows the change in the destruction term of the Boltzmann equation for protons as a function of $\mu_1 =\cos \theta_1$, i.e., $\Delta df/dt(\mu_1) /[ (1/2) \int d\mu_1 df/dt(\mu_1)]$, at $T=10^9$ K for $B =2.0\times 10^{11}$ G. The thin three lines near zero are differences in the  total destruction rates. It is confirmed that the proton distribution function is independent of the direction even if the isotropy of the $e^\pm$ momentum is broken in a  magnetic field. Also plotted are thick lines for the partial destruction rate by $e^\pm$ particles moving toward the hemisphere of $q_{e\parallel}>0$. At high energies relevant to nuclear reactions, a significant dependence on $\mu_1$ is seen (thick solid line for $E_A/T=20$). The partial destruction rate is smaller at larger $\mu_1$ since the electron momentum in the nuclear rest frame is smaller than that in the fluid rest frame. However, after integration over the full phase space of $\bfq_e$, the dependence of the destruction rate on $\mu_1$ diminishes (thin horizontal lines).

The lower panel in Fig.~\ref{fig2} shows the destruction term averaged over the proton direction, i.e., $\langle df/dt\rangle_{\mu_1} =\int d\mu_1 df/dt(\mu_1) /2$, as a function of the proton energy. Solid and dotted lines correspond to the results for $B=2.0 \times 10^{11}$ G and 0, respectively. There is no significant difference, and it is also confirmed that the magnetic field does not affect the proton energy distribution. Note that, although the total $e^\pm$ number density is different from that of a MB distribution in a magnetic field \cite{Kernan:1995bz}, the difference is small for the upper bound on the field strength adopted here.

%*******************************************************

%*******************************************************

\section{Constraint on anisotropy}\label{sec4}

We derive a constraint on anisotropy in the nuclear momentum distribution during BBN. In this case, modifications to the MB distribution can be formulated as a change in the effective relative velocity distribution. Similarly to a deviation of the energy spectrum \cite{Bertulani:2012sv,Hou:2017uap}, an introduction of anisotropy affects nucleosynthesis. As an example, we assume that the densities of particles moving in some special direction are enhanced, as described by $f_{\rm a}(\bfv_i) \propto g_{\rm a}(\mu_i) \equiv (1+a\mu_i)$ (for $\mu_i \equiv \bfv_i \cdot \bfB /(v_i B)>-1/a$) and 0 (otherwise). Similar to Ref. \cite{Kusakabe:2018dzx}, the relative velocity distribution function can then be written as
\begin{eqnarray}
  f_{\rm a}^{\rm rel}(\bfv) &=&
  \int_0^{2 \pi} d\phi_{\rm CM} \int_{-1}^1 d\mu_{\rm CM} \int_0^\infty V^2 dV
  f_{\rm a}(\bfv_A) f_{\rm a}(\bfv_e) \nonumber \\
  &=&
  B_1(a)^2
  \frac{\left( m_1 m_2 \right)^{3/2}}{\left( 2 \pi T \right)^3}
  \exp\left[ -\frac{\mu v^2}{2 T} \right]
  \nonumber \\
  && \times
  \int_0^\infty V^2 dV \exp\left[ -\frac{M V^2}{2 T} \right]
  \int_0^{2 \pi} d\phi_{\rm CM}
  \nonumber \\
  && \times
  \int_{-1}^1 d\mu_{\rm CM} g_{\rm a}(\mu_1) g_{\rm a}(\mu_2),
\end{eqnarray}
where $B_1(0 \leq a \leq 1) =1$ and $B_1(a> 1) =2 /[1 +a/2 +1 /(2a)]$ are the normalization constants,
$m_1$ and $m_2$ are masses of reacting particles 1 and 2,
$M=m_1 +m_2$,
$V$ is the center of mass (CM) velocity, while 
$\mu_{\rm CM}$ and $\phi_{\rm CM}$ are, respectively, the cosine of the polar angle and azimuthal angle at the scattering.

\begin{figure}[tbp]
\begin{center}
\includegraphics[width=\linewidth]{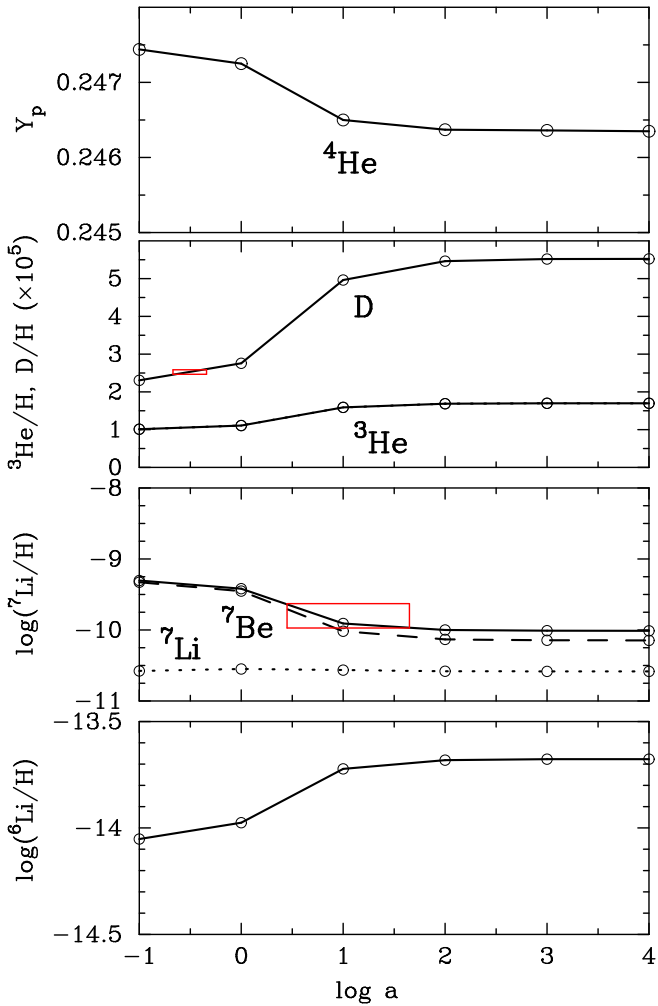}
\caption{Primordial abundances as a function of the anisotropy parameter $a$ (open circles). $Y_p$ is the mass fraction of $^4$He. Solid lines connect the calculated total abundances, while the dashed and dotted lines correspond to separated yields of $^7$Be and $^7$Li in  BBN, respectively. Boxes roughly indicate values of the anisotropy parameter where the observational constraints  are satisfied.
  \label{fig3}}
\end{center}
\end{figure}

Figure \ref{fig3} shows light element abundances as a function of the anisotropy parameter $a$ in the BBN model including the anisotropy in all nuclear distribution functions.
 The introduction of this type of anisotropy causes larger energies of the CM, i.e., $MV^2/2$, and smaller relative energies in the CM system, i.e., $E=\mu v^2/2$. As a result, a larger $a$ leads to a softer distribution of $E$ with a smaller average. Then, nuclear reaction rates are hindered and light element abundances are changed in a similar  manner as in the case where the isotropic distribution function  deviates from a MB distribution to the softer side. The dependencies of abundances on anisotropy is very similar to those of a spectral distortion \cite{Hou:2017uap,Kusakabe:2018dzx}.

%*******************************************************

%*******************************************************

\section{Summary}\label{sec5}
The effect of Landau discretization of the $e^\pm$ momenta on the collision term of protons during BBN was studied. At $T=10^9$ K, nuclei experience $\sim 10^{21}$ scatterings per second from $e^\pm$, and most of the scatterings lead to small, $\mathcal{O}$(1) eV, changes of the nuclear kinetic energy.
When the magnetic field amplitude is set to the largest value allowed by observations, the $e^\pm$ distribution experiencing the magnetic field does not lead to any significant directional dependence in the collisional destruction term of nuclei.
A constraint on the amplitude of anisotropies in the nuclear distribution function during BBN was also derived.  

\begin{acknowledgments}
  This work is supported by NSFC Research Fund for International Young Scientists (Grant No. 11850410441).
   Work at the University of Notre Dame supported by DOE nuclear theory grant DE-FG02-95-ER40934.
\end{acknowledgments}

\bibliography{reference3}

%%%%%%%%%%%%%%%%%%%%%%%%%%%%%%%%%%%%%%%%%%%%%%%%%%%%%%
%Merlin.mbs v4.21 2009-07-09.
%\begin{thebibliography}{99}%

%\end{thebibliography}
\end{document}